\documentclass[multphys,vecphys]{svmult}

\usepackage{makeidx}         
\usepackage{graphicx}        
\usepackage{multicol}        
\usepackage[bottom]{footmisc}

\makeindex             

\newcommand{\simgt}{\lower.5ex\hbox{$\; \buildrel > \over \sim \;$}}
\newcommand{\simlt}{\lower.5ex\hbox{$\; \buildrel < \over \sim \;$}}

\begin{document}

\title*{Observational Constraints on the ICM Temperature Enhancement by Cluster Mergers
}
\titlerunning{Constrains on Temperature Enhancement by Cluster Mergers}
\author{N. Okabe\inst{1} \and K. Umetsu\inst{2}}
\institute{Astronomical Institute, Tohoku University, Aramaki, Aoba-ku, Sendai, 980-8578, Japan
\texttt{okabe@astr.tohoku.ac.jp}
\and Academia Sinica Institute of Astronomy and Astrophysics (ASIAA) P.O. Box 23-141, Taipei 106, Taiwan,
R.O.C. \texttt{keiichi@asiaa.sinica.edu.tw}}

%
%
\maketitle
\vspace{-0.2cm}
\section*{Abstract}
\vspace{-0.2cm}
We present results from a combined weak lensing and X-ray analysis
of the merging cluster A1914\index{A1914} at a redshift of $z=0.1712$
based on $R_{\rm}$-band imaging data with Subaru/Suprime-Cam
and archival Chandra X-ray data.
Based on the weak-lensing and X-ray data
we explore 
the relationships between cluster global properties, namely
the gravitational mass, 
the bolometric X-lay luminosity and temperature, the gas mass 
and the gas mass fraction, as a function of radius.
%
We found that the gas mass fractions
within $r_{2500}$ and $r_{{\rm vir}}$ are 
consistent with the results of earlier X-ray cluster studies
and of cosmic microwave background studies based on the WMAP
observations, respectively.
%
However, the observed temperature, $k_B  T_{\rm ave}=9.6\pm0.3 {\rm
keV}$, 
is significantly higher than the virial temperature, 
$k_B T_{\rm vir}=4.7\pm0.3 {\rm keV}$ derived from the weak lensing
distortion measurement.
The X-ray bolometric luminosity and temperature
 ($L_X-T$) relation is consistent with the $L_X-T$ relation derived by previous
 statistical X-ray studies of galaxy clusters.
Such correlations among the global cluster properties are invaluable
observational tools for studying the cluster merger physics.
Our results demonstrate that the combination of X-ray and weak-lensing
observations is a promising, powerful probe of the physical processes 
associated with cluster mergers as well as of their mass properties.

\vspace{-0.5cm}

\section{Weak Lensing and X-ray Analyses}
\label{sec:1}
\vspace{-0.1cm}
\subsubsection*{Weak Lensing Analysis}
\vspace{-0.2cm}
Weak lensing provides a direct measure of the projected mass
distribution in the universe regardless of the physical/dynamical state
of matter in the system.
Therefore weak lensing enables the direct study of mass in clusters
even when the clusters are in the process of (pre/mid/post) merging,
where the assumptions of hydrostatic equilibrium or isothermality 
are no longer valid.
We carried out a weak lensing analysis on the merging cluster A1914
\index{A1914} with deep $R_{\rm c}$-band data taken with Suprime-Cam
on the Subaru telescope,
covering the entire cluster region out to the cluster virial radius
thanks to the wide field-of-view of $34'\times 27'$.
The details of the analysis will be presented in
Okabe \& Umetsu \cite{ou07} and Umetsu \& Okabe\cite{uo07}. 
We define a sample of background galaxies with 
magnitudes
$ 21 \simlt R_{\rm c} \simlt 26$ 
and $r_h$
and half-light radii 
$r_h^* \simlt r_h \simlt 15 {\rm pixels}$,
yielding a mean galaxy number density of 
$n_g\simeq 48 {\rm arcmin}^{-2}$.
%
We derived a radial profile of the tangential component of reduced
gravitational shear, $g_+=\gamma_+/(1-\kappa)$,
over the radial range of $3'$ to $17'$,
where the irregularity in the mass distribution 
is less significant than that in the central region.
The best-fitting NFW (Navarro, Frenk, \& White\cite{nfw97})
profile to the Subaru distortion data 
is obtained as follows:
virial mass $M_{{\rm vir}}=(7.66\pm0.73)\times 10^{14}
 M_{\odot} h_{70}^{-1}$
(or viral radius $r_{\rm vir}=12'.26=2.144 h_{70}^{-1}{\rm Mpc}$);
concentration parameter $c=5.07\pm1.75$. 
The best-fitting mass profile from weak lensing distortion measurements
is in good agreement with
the luminosity profile of cluster member galaxies,
which will be presented in Umetsu \& Okabe \cite{uo07}.

\vspace{-0.5cm}

\begin{figure}
\hspace{5mm}
\begin{minipage}{.35\linewidth}
\centering
\includegraphics[height=3.5cm]{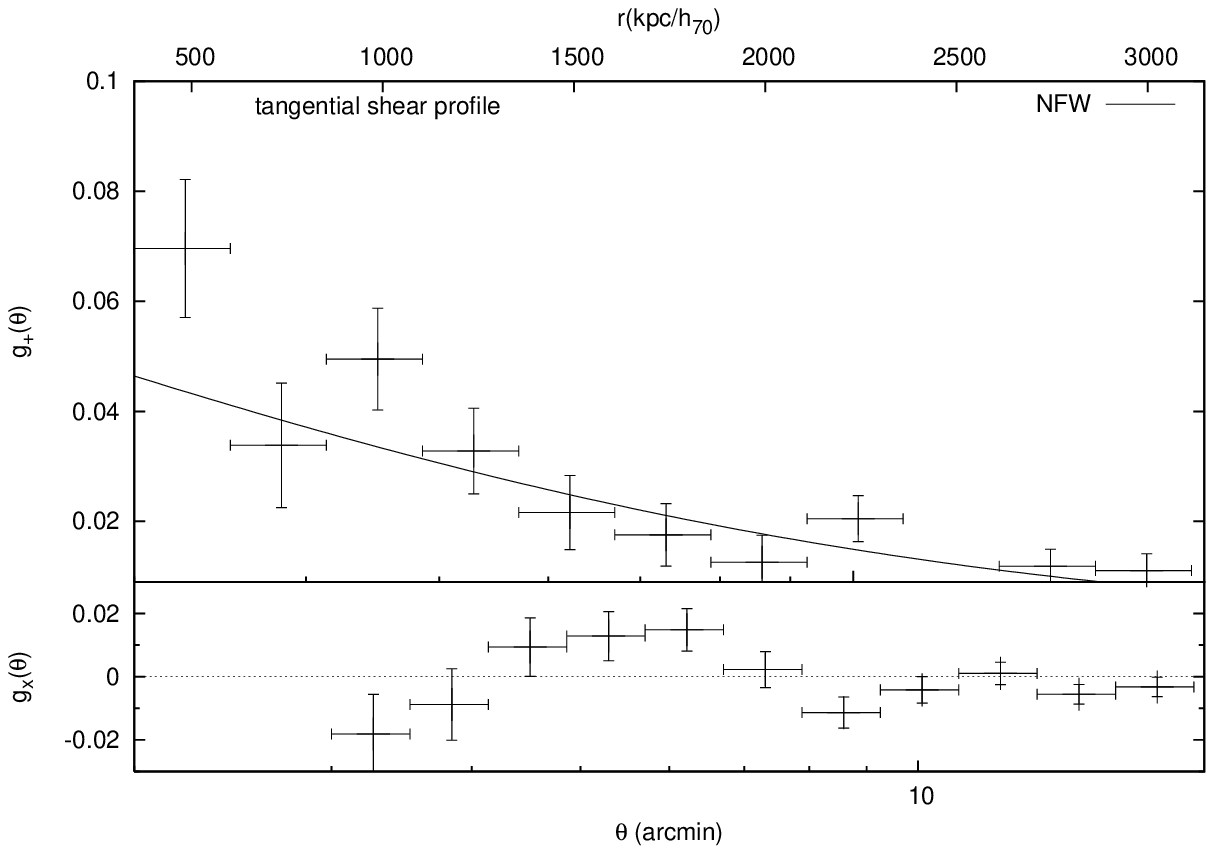}
\caption{\scriptsize{
Radial profiles of the reduced tangential shear (upper panel)
 and the $45^\circ$ rotated ($\times$) component (lower panel).
The solid curve represents the best-fitting  NFW shear profile.}}
\label{fig:shear}       
\end{minipage}
\hspace{20mm}
\begin{minipage}{.35\linewidth}
\includegraphics[height=3.5cm]{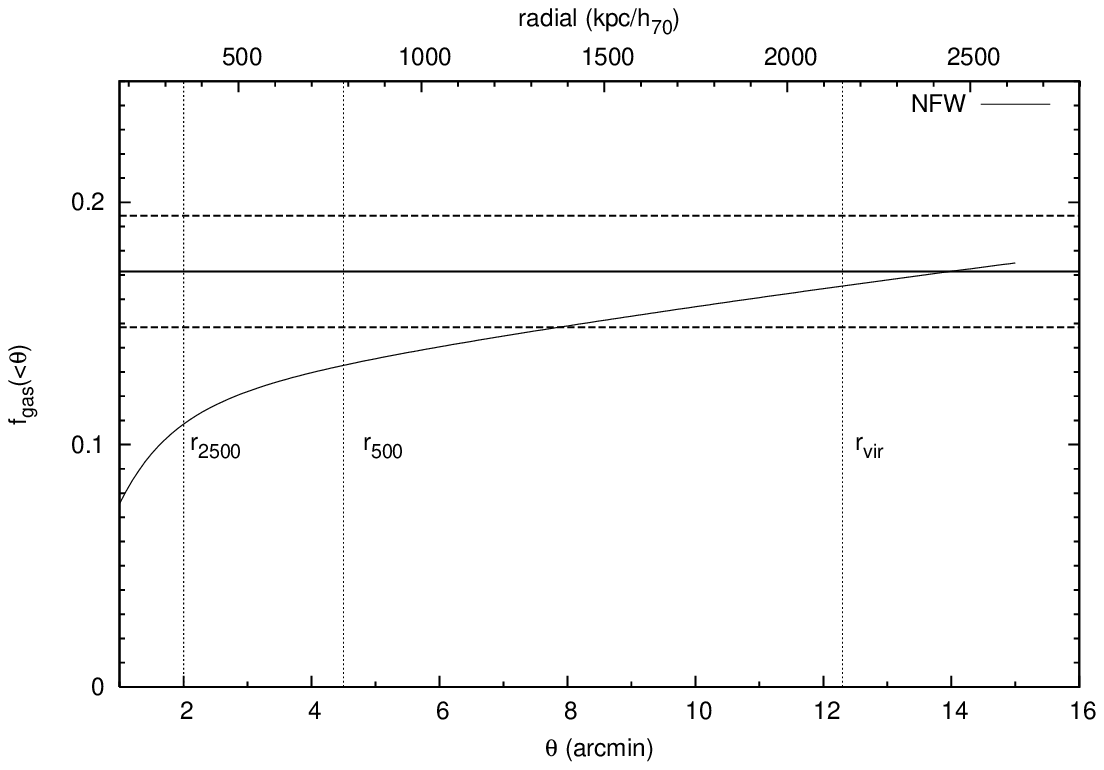}
\caption{\scriptsize{Radial profile  of 
the cluster gas mass fraction, $f_{\rm gas}$, 
obtained by combining best fit results of weak lensing and X-ray analyses.
Dashed vertical lines indicate the $1\sigma$ confidence range of
the cosmic baryon fraction.}}\label{fig:fgas}  
\end{minipage}
\end{figure}


\vspace{-1.2cm}
\subsubsection*{X-ray Analysis}
\vspace{-0.2cm}
We used archival {\it Chandra} data to 
measure physical properties of the ICM in A1914.
We performed a spectral fit
with a single temperature model within a radius of $5'$.
The average temperature and abundance were obtained as
$k_B T_{{\rm ave}}(\theta<5')=9.6\pm0.3 {\rm keV}$
$A=0.19\pm0.08$ at a $90\%$ confidence level, respectively. 
The radial profile fit was performed on the observed X-ray surface
brightness distribution ($\theta<6'$) using a single $\beta$ model,
where we adopted the same center as for the weak lensing tangential
shear measurement.
The best fitting parameters were obtained as follows:
cluster core radius, $r_c=1.'03\pm0.'06$;
slope parameter, $\beta=0.727\pm0.014$; 
central electron density,
$n_{e,0}=(1.46\pm0.26)\times10^{-2} h_{70}^{-2} {\rm cm}^{-3}$.




\vspace{-0.5cm}

\section{Global Parameters of Merging Clusters}
\label{sec:1}
\vspace{-0.1cm}
\subsubsection*{Gas Mass Fraction}
\vspace{-0.2cm} 
We show in Figure \ref{fig:fgas}
the radial profile of the gas mass fraction, 
$f_{\rm gas}(<r)=M_{\rm gas}(<r)/M_{\rm tot}(<r)$.
We used the best-fitting models derived in \S1 to calculate
$f_{\rm gas}(<r)$.
Using the $\beta$ model we extrapolated the gas mass profile outside
$\theta=6'$ where the weak lensing mass profile is available.
The values within typical radii of $r_{2500}$, $r_{500}$ and 
$r_{{\rm vir}}$, 
at which the mean density is
$2500$, $500$ and $\delta_{\rm vir}$ times the critical density of the
 universe, are $f_{2500}=0.108$, 
$f_{500}=0.133$ and $f_{{\rm vir}}=0.165$, respectively. 
The central gas mass fraction,
$f_{2500}$, is similar to those values for relaxed
 clusters derived by X-ray data alone, 
such as $0.091\pm0.002$
 (Vikhlinin et al. \cite{vik06}) and $0.117\pm0.002$ 
(Allen et al. \cite{all04}).
The values of $f_{500}$ deduced from X-ray studies 
show a large scatter among different clutters and different observations. 
We note the values of $f_{500}$ deduced from X-ray studies
show a large scatter among different clutters and different observations.
We found the virial gas fraction,
$f_{\rm vir}$, of A1914 is consistent with the cosmic mean baryon fraction
$\Omega_{b}/\Omega_{m}$, constrained by the CMB observations (Spergel et
al. \cite{spe03}).
The virial gas fraction,
$f_{\rm vir}$, is consistent with the cosmic mean baryon fraction
$\Omega_{b}/\Omega_{m}$, constrained by the CMB observations
(Spergel  et al. \cite{spe03}). 
We emphasize that 
a combined weak-lensing and X-ray study will allow us to 
determine gas mass fractions even in merging clusters.




\vspace{-0.5cm}
\subsubsection*{$L_X-T$ relation}
\vspace{-0.2cm}
We derived the X-ray bolometric luminosity $L_{X,{\rm bol}}$
and temperature $T$
with a  single-temperature model. 
The resulting X-ray luminosity within
$r_{200}$ is 
$L_{X,{\rm bol}}=2.5 \times10^{45} h_{70}^{-2}{\rm ergs~s}^{-1}$
with $k_B T_{\rm vir}=4.7^{+0.3}_{-0.3}\left(M/M_{\rm
vir}\right)^{2/3} {\rm keV}$,
which is consistent with the observed local $L_X-T$ relation
from previous X-ray studies:
$L_{\rm X,bol}(<r_{200})=2.3^{+0.3}_{-0.3}\times10^{45}\left(k_B
T/9.6{\rm keV}\right)^{2.88\pm0.15} h_{70}^{-2}$ ${\rm ergs~s}^{-1}$
(Arnaud \& Evrard \cite{arn99}). 
This indicates that the observed local $L_X-T$ relation holds
even in the merging cluster A1914.

\vspace{-0.5cm}

\subsubsection*{$M-T$ relation}
\vspace{-0.2cm}
The observed temperature $k_B T=9.6\pm0.3 {\rm keV}$
is significantly higher than the viral
temperature, $k_B T_{\rm vir}=4.7^{+0.3}_{-0.3}\left(M/M_{\rm
vir}\right)^{2/3} {\rm keV}$ 
derived from the weak lensing analysis.
If the cluster were virialized before the merging process, 
then the ICM temperature could be heated up by a factor of two. 
Based on this scenario, we
constrain the heating energy of the ICM induced by the 
cluster merger as\vspace{-0.3cm}
\begin{eqnarray}
\langle \Delta E_{{\rm ICM}}(r<5') \rangle \rangle &=& 4 \pi (k_B T_{{\rm
  ave}}- k_B T_{{\rm vir}}) \int \sum_j n_j(r) r^2 dr \sim 2 \times 10^{62} {\rm erg}, \nonumber
\end{eqnarray} 
\vspace{-0.5cm}

with $n_H=0.82n_e$, where we assumed the electron and ion
temperatures are the same. 
\vspace{-0.8cm}
%
%

%
%



\printindex
\end{document}